\documentclass[preprint,showpacs]{revtex4}

\usepackage{graphicx}% Include figure files
\usepackage{dcolumn}% Align table columns on decimal point
\usepackage{bm}% bold math

\begin{document}

\title{Asymmetric nuclear matter in the relativistic mean field
       approach with vector cross-interaction}

\author{Juraj Kotuli\v{c} Bunta}
\email{juraj.bunta@savba.sk}

\author{\v{S}tefan Gmuca}
\email{gmuca@savba.sk}

\affiliation{Institute of Physics, Slovak Academy of Sciences,
Dubravska cesta 9, SK-845 11 Bratislava 45, Slovakia}

\date{\today}

\begin{abstract}
Asymmetric nuclear matter is studied in the frame of relativistic
mean-field theory, using scalar-isoscalar $\sigma$,
vector-isoscalar $\omega$ meson together with their
selfinteractions, vector-isovector $\rho$ meson with its
cross-interaction with $\omega$ meson too, and scalar-isovector
$\delta$ meson as degrees of freedom. The model is used to
parameterize the nuclear matter properties results calculated by
more fundamental Dirac-Brueckner-Hartree-Fock theory and thus to
provide an effective DBHF model applicable also to finite nuclei.
Vector $\omega$-$\rho$ cross-interaction seems to be an useful
degree of freedom for describing of the asymmetric nuclear matter,
mostly due to its impact on density dependence of the symmetry
energy.
\end{abstract}



\maketitle

\section{Introduction}
\label{sec:Intro}

The study of nuclear matter -- hypothetical uniform infinite
system of nucleons interacting via strong forces -- has already
been an important part of the nuclear physics development for
several decades. It is a good starting point for both, nuclear
physics of finite systems (e.g. structure and properties of finite
nuclei, dynamics of heavy-ion collisions) and astrophysics (e.g.
structure and evolution of stars). The early attempts were based
on the nonrelativistic Brueckner--Hartree--Fock (BHF) theory (see
e.g. Refs. \cite{BHF} for review and references therein), and
related the bare nucleon--nucleon (NN) interaction to nuclear
ground--state properties in a parameter--free way with a limited
success.

The break--through was achieved when the relativistic extension of
the BHF theory (so called Dirac--Brueckner--Hartree--Fock (DBHF)
approach) was developed \cite{Shakin,BM84,HM87} and successfully
applied to nuclear matter problems. An essential feature of the
DBHF is the incorporation of the relativistic dynamics, governed
by the Dirac equation with strong scalar and vector fields. The
explicit treatment of the lower components of Dirac spinors gives
a rise to strongly density--dependent relativistic effects. They
shift the nuclear matter saturation points (``Coester band'')
towards empirical values \cite{BM90}. Subsequently, a great effort
has been devoted to the successful DBHF description of both,
symmetric and (to lesser extent) isospin--asymmetric nuclear
matter.

Thus, the DBHF approach is currently considered to be a
microscopic parameter--free nuclear model based on realistic NN
interaction. However, due to its complexity, this sophisticated
approach is successfully manageable for nuclear matter properties
only; finite nuclei are at present beyond the scope of this model.
To overcome this restriction, several approaches were developed
which relate the DBHF output for nuclear matter to the parameters
of the relativistic mean--field (RMF) theory \cite{SW86,SW97}. The
RMF approach is widely used and powerful phenomenological tool for
various aspects of nuclear many--body problems which provides an
effective framework for calculation of both, nuclear matter and
(contrary to DBHF) also finite nuclei. The nonlinear RMF approach
\cite{BB77} has already been proven to be a reliable tool for the
calculation of normal nuclei close to the valley of stability
\cite{GRT90}, exotic and superheavy nuclei \cite{BR99}.

Now new experimental facilities are available to study properties
of exotic nuclei with high isospin asymmetry. Additionally,
increasingly more precise observations and measurements of
properties of neutron stars and supernovae have been carried out.
This naturally brings a need for better description of isospin
degree of freedom, which can be done by enhancing the isovector
meson sector. The isovector scalar $\delta$ meson \cite{KK97} and
vector cross-interactions \cite{MS96} were included into RMF for
this purpose.

The goal of this paper is obtaining of effective parametrization
applied to asymmetric nuclear matter, using different degrees of
freedom in order to study influence of $\delta$ meson and vector
meson cross-interaction on quality of the reproduction of DBHF
results, as well as their influence on calculated nuclear matter
properties, especially on density dependence of nuclear symmetry
energy.

\section{Theoretical framework}

The starting point of the model is Lagrangian density that
introduces nucleon field $\psi$, isoscalar-scalar meson field
$\sigma$, isoscalar-vector meson field $\bm{\omega}$,
isovector-vector meson field $\bm{\rho}$ and isovector-scalar
meson field $\delta$ (pion field does not contribute, because it
is pseudoscalar, and nuclear matter is parity invariant), and
takes a form

\begin{center}
\begin{eqnarray}
&{\cal{L}}(\psi,\sigma,\bm{\omega},\bm{\rho},\delta)= \bar{\psi}
\left[\bm{\gamma}_{\mu}(i\partial^{\mu}-g_{\omega}\bm{\omega}^{\mu}
-(M-g_{\sigma}\sigma)\right]\psi
\nonumber\\
&+\frac{1}{2}(\partial_{\mu}\sigma\partial^{\mu}\sigma
-m_{\sigma}^{2}\sigma^{2})
-\frac{1}{4}\bm{\omega}_{\mu\nu}\bm{\omega}^{\mu\nu}
+\frac{1}{2}m_{\omega}^{2}\bm{\omega}_{\mu}
\bm{\omega}^{\mu}
\nonumber\\
\label{lagrangian}
&-\frac{1}{3}b_{\sigma}M{(g_{\sigma}\sigma)}^{3}-\frac{1}{4}c_{\sigma}
{(g_{\sigma}\sigma)}^{4}
+\frac{1}{4}c_{\omega}{(g_{\omega}^{2}
\bm{\omega}_{\mu}\bm{\omega}^{\mu})}^{2}
\nonumber\\
&+\frac{1}{2}(\partial_{\mu}\delta\partial^{\mu}\delta
-m_{\delta}^{2}{\delta}^{2})
+\frac{1}{2}m_{\rho}^{2}\bm{\rho}_{\mu}.\bm{\rho}^{\mu}
-\frac{1}{4}\bm{\rho}_{\mu\nu}.\bm{\rho}^{\mu\nu}
\nonumber\\
&+\frac{1}{2}\Lambda_{V}(g_{\rho}^{2}\bm{\rho}_{\mu}.\bm{\rho}^{\mu})
(g_{\omega}^{2}\bm{\omega}_{\mu}\bm{\omega}^{\mu})
\nonumber\\
&-g_{\rho}\bm{\rho}_{\mu}\bar{\psi}\gamma^{\mu}\bm{\tau}\psi
+g_{\delta}\delta\bar{\psi}\bm{\tau}\psi\;,
\end{eqnarray}
\end{center}
where antisymmetric field tensors are given by

\begin{eqnarray*}
\bm{\omega}_{\mu\nu}\equiv\partial_{\nu}\bm{\omega}_{\mu}
-\partial_{\mu}\bm{\omega}_{\nu}
\;,
\\
\bm{\rho}_{\mu\nu}\equiv\partial_{\nu}\bm{\rho}_{\mu}
-\partial_{\mu}\bm{\rho}_{\nu}\;,
\end{eqnarray*}
and the symbols used have their usual meaning. The parameters
entering the lagrangian are $M$ that denotes the nucleon free
mass, whereas $m_{\sigma}$, $m_{\omega}$, $m_{\varrho}$ and
$m_{\delta}$ are masses assigned to the meson fields. The first
term together with the last two ones describe interaction of
isoscalar and isovector mesons with nucleons where the strength of
these interactions is determined by dimensionless coupling
constants $g_{\sigma}$, $g_{\omega}$, $g_{\varrho}$ and
$g_{\delta}$. Three terms in the third line represent cubic and
quartic scalar selfinteractions \cite{BB77} and quartic vector
selfcouplings \cite{Bodmer91,Gmuca91}, the strength of which is
also given by dimensionless selfinteraction coupling constants
$b_{\sigma}$, $c_{\sigma}$ and $c_{\omega}$. The second and fourth
lines represent free (noninteracting) lagrangian for all mesons,
and the fifth line realizes cross-interaction between $\omega$ and
$\rho$ mesons characterized by cross-coupling constant
$\Lambda_{V}$ \cite{MS96}.

Constraint of stationarity of the action leads to the well-known
Euler-Lagrange field equations and equations of motion follow
after their application to the lagrangian (\ref{lagrangian}). This
produces the Dirac equation for nucleon field

\begin{equation} \label{dirac}
\left[\bm{\gamma}_{\mu}(i\partial^{\mu}-g_{\omega}\bm{\omega}^{\mu}
-g_{\rho}\bm{\rho}_{\mu}.\bm{\tau}) -
(M-g_{\sigma}\sigma-g_{\delta}\bm{\delta}.\bm{\tau})\right]\psi=0\;.
\end{equation}

Isoscalar meson field $\sigma$, $\omega$ are then described by
Klein-Gordon and Proca equations, respectively,

\begin{equation}
\label{sigmamotion}
(\partial_{\mu}\partial^{\mu}+m_{\sigma}^{2})\sigma=
g_{\sigma}[\bar{\psi}\psi - b_{\sigma}M{(g_{\sigma}\sigma)}^{2}
-c_{\sigma}{(g_{\sigma}\sigma)}^{3}]\;,
\end{equation}
\begin{eqnarray}
\label{omegamotion}
\partial_{\mu}\bm{\omega}^{\mu\nu}+m_{\omega}^{2}\bm{\omega}^{\nu}&=&
g_{\omega}[\bar{\psi}\bm{\gamma}^{\nu}\psi-c_{\omega}g_{\omega}^{3}
(\bm{\omega}_{\mu}\bm{\omega}^{\mu}\bm{\omega}^{\nu})
\nonumber\\
&&-\Lambda_{V}g_{\rho}^{2}\bm{\rho}_{\mu}.\bm{\rho}^{\mu}
 g_{\omega}\bm{\omega}_{\mu}]\;.
\end{eqnarray}

Analogically, isovector $\rho$ and $\delta$ meson fields read,

\begin{equation}
\label{rhomotion}
\partial_{\mu}\bm{\rho}^{\mu\nu}+m_{\rho}^{2}\bm{\rho}^{\nu}=
g_{\rho}[\bar{\psi}\bm{\gamma}^{\mu}\bm{\tau}\psi
-\Lambda_{V}g_{\rho}\bm{\rho}_{\mu}
g_{\omega}^{2}\bm{\omega}_{\mu}\bm{\omega}^{\mu}]\;,
\end{equation}
\begin{eqnarray}
\label{deltamotion}
(\partial_{\mu}\partial^{\mu}+m_{\delta}^{2})\delta&=&
g_{\delta}\bar{\psi}\bm{\tau}\psi\;.
\end{eqnarray}

Due to the fact that these equations are nonlinear, nowadays there
is known no suitable method to solve them. The way to avoid this
is to replace the operators of meson fields by their expectation
values -- the so called mean-field approximation. The fields are
thus treated as classical c-numbers. Its reasonability increases
with increasing baryon density. The second approximation
introduced is the non-sea approximation which doesn't take account
of the Dirac sea of negative energy states.

In this model we are dealing with static, homogenous, infinite
nuclear matter that allows us to consider some other
simplifications due to translational invariance and rotational
symmetry of nuclear matter. This causes the expectation values of
space-like components of vector fields vanish and only zero
components - $\rho_{0}$ and $\omega_{0}$ - survive \cite{GRT90}.
In addition, rotational invariance around third axe of isospin
space results in taking into account only the third component of
isovector fields - $\rho^{(3)}$ and $\delta^{(3)}$
\cite{SW97,KK97}. The above-mentioned can formally be written as

\begin{eqnarray*}
\sigma &\longrightarrow&
\left\langle\sigma\right\rangle\equiv\bar{\sigma}\;,
\\
\bm{\omega}_{\mu} &\longrightarrow&
\left\langle\bm{\omega}_{\mu}\right\rangle \equiv\delta_{\mu
0}\bm{\bar{\omega}}_{\mu}=\bar{\omega}_{0}\;,
\\
\bm{\rho}_{\mu}&\longrightarrow&\left\langle\bm{\rho}_{\mu}\right\rangle
\equiv\bar{\rho}_{0}^{(3)}\;,
\\
\delta&\longrightarrow&\left\langle\delta\right\rangle\equiv
\bar{\delta}^{(3)}\;.
\end{eqnarray*}

Having inserted the above simplifications the field equations are
reduced and we can easily obtain potentials of both isoscalar
meson fields

\begin{eqnarray}
\label{sigmapotential} U_{\sigma}&\equiv&
-g_{\sigma}\bar{\sigma}=-\frac{g_{\sigma}^{2}}{m_{\sigma}^{2}}
[\rho_{S}- b_{\sigma}M{(g_{\sigma}\bar{\sigma})}^{2}
\nonumber\\
&&-c_{\sigma}{(g_{\sigma}\bar{\sigma})}^{3}] \;,
\\
\label{omegapotential} U_{\omega}&\equiv&
g_{\omega}\bar{\omega}_{0}=\frac{g_{\omega}^{2}}{m_{\omega}^{2}}
[\rho_{B}-c_{\omega}{(g_{\omega}\bar{\omega}_{0})}^{3}
\nonumber\\
&&-U_{\rho}^{2}\Lambda_{V}(g_{\omega}\bar{\omega}_{0})]\;,
\end{eqnarray}
and isovector meson fields

\begin{eqnarray}
\label{rhopotential} U_{\rho}&\equiv&
g_{\rho}\bar{\rho}_{0}^{(3)}= \frac{g_{\rho}^{2}}{m_{\rho}^{2}}[
\bar{\psi}\gamma^{0}\tau_{3}\psi-g_{\rho}\bar{\rho}_{0}^{(3)}
\Lambda_{V}{(g_{\omega}\bar{\omega}_{0})}^{2}]=
\nonumber\\
&&\frac{g_{\rho}^{2}}{m_{\rho}^{2}}[(2\frac{Z}{A}-1)\rho_{B}
-g_{\rho}\bar{\rho}_{0}^{(3)}
\Lambda_{V}U_{\omega}^{2}]\;,
\\
\label{deltapotential} U_{\delta}&\equiv&
-g_{\delta}{\bar{\delta}}^{(3)}=
-\frac{g_{\delta}^{2}}{m_{\delta}^{2}}\bar{\psi}\tau_{3}\psi=
\frac{g_{\delta}^{2}}{m_{\delta}^{2}}(\rho_{n}^{S}-\rho_{p}^{S})\;,
\end{eqnarray}
where scalar density $\rho_{S}$ is expressed as the sum of proton
(p) and neutron (n) part

\begin{equation} \label{densityscalar}
\rho_{S}=\left\langle\bar{\psi}\psi\right\rangle=\rho_{p}^{S}
+\rho_{n}^{S}\;,
\end{equation}
which are given by

\begin{equation} \label{densityscalarpn}
\rho_{i}^{S}=\frac{2}{{(2\pi)}^{3}}\int_{0}^{k_{i}}d^{3}\!k \,
\frac{M_{i}^{*}}{{(\bm{k}^{2}+{M_{i}^{*}}^{2})}^{1/2}}\;, i=p,n
\;.
\end{equation}

In the equation (\ref{densityscalarpn}) $k_{i}$ is nucleons' Fermi
momentum and $M_{p}^{*}$, $M_{n}^{*}$ denotes proton and neutron
effective masses, respectively, which can be written as

\label{nucleonmasses}
\begin{eqnarray}
M_{p}^{*}=M-g_{\sigma}\bar{\sigma}-g_{\delta}{\bar{\delta}}^{(3)}\;,
\\
M_{n}^{*}=M-g_{\sigma}\bar{\sigma}+g_{\delta}{\bar{\delta}}^{(3)}\;.
\end{eqnarray}

One can see that condensated scalar $\sigma$ meson field generates
a shift of nucleon mass, in consequence of which nuclear matter is
described as a system of pseudonucleons with masses $M^{*}$ moving
in classical fields $\bar{\sigma}$, $\bar{\omega}_{0}$ and
${\bar{\rho_{0}}}^{(3)}$, where additionally $\delta$ meson field
is responsible for splitting of proton and neutron effective
masses, which is an important feature of $\delta$ meson influence
on the nuclear matter saturation mechanism and its properties. The
$\delta$ meson seemed to be an useful degree of freedom in
describing of asymmetric nuclear matter, indicated by its
influence on e.g. stiffness of equation of state, slope and
curvature of symmetry energy and properties of warm asymmetric
nuclear matter \cite{KK97,LGBC02}.

The solution requires to be performed selfconsistently, which can
be clearly seen from equations
(\ref{sigmapotential})-(\ref{deltapotential}) where $\sigma$
potential (\ref{sigmapotential}) must be solved using iterations.

The baryon density is given by

\begin{equation} \label{densitybaryon}
\rho_{B}=\left\langle\bar{\psi}\gamma^{0}\psi\right\rangle=
\frac{4}{{(2\pi)}^{3}}\int_{0}^{k_{F}}d^{3}\!k=
\frac{2}{3\pi^{2}}k_{F}^{3}\;,
\end{equation}
with $k_{F}$ beeing an average Fermi momentum. It can be seen that
scalar density (\ref{densityscalar}) is less than baryon density
due to term $M_{i}^{*}/{(\bm{k}^{2}+{M_{i}^{*}}^{2})}^{1/2}$ that
causes reduction of the contribution of rapidly moving nucleons to
scalar source term. This mechanism is responsible for nuclear
matter saturation in the mean field theory and essentially
distinguishes relativistic models from nonrelativistic ones.

Cross-coupling of the $\omega$ and $\rho$ mesons requires also
selfconsistent calculation of equations (\ref{omegapotential}) and
(\ref{rhopotential}), with iterative procedure for $\omega$
potential.

By reason that $\delta$ field splits nucleon effective masses, the
proton and neutron Fermi momenta will be also splitted, while they
have to fulfil

\begin{equation}
\label{densitysum} \rho_{B}=\rho_{p}+\rho_{n}=
\frac{2}{{(2\pi)}^{3}}\int_{0}^{k_{p}}d^{3}\!k+
\frac{2}{{(2\pi)}^{3}}\int_{0}^{k_{n}}d^{3}\!k\;,
\end{equation}
where $k_{F}$ is average Fermi moment of the matter, $k_{p}$,
$k_{n}$ are Fermi momenta of protons and neutrons, respectively.
The different values of Fermi momenta have consequences for
transport properties of asymmetric nuclear matter.

To obtain formula for energy density of nuclear matter it is
essential to have cognizance of the energy tensor, in continuum
mechanics defined as \cite{FW80}

\begin{equation} \label{energytensor}
T_{\mu\nu}=-g_{\mu\nu}\mathcal{L}+\frac{\partial\Phi_{i}}{\partial
\bm{x}^{\nu}}
\frac{\partial\mathcal{L}}{\partial(\partial\Phi_{i}/\partial
\bm{x}_{\mu})}\;,
\end{equation}
where $\Phi_{i}$ generally denotes physical fields. The energy
density of such a system is the zero component of the energy
tensor $\varepsilon=\left\langle T_{00}\right\rangle$, and finally
the binding energy per nucleon is related to energy density by

\begin{equation} \label{energypernucleon}
E_{b}=\frac{\varepsilon}{\rho_{B}}-M\;.
\end{equation}

The energy density per nucleon is a starting quantity for further
properties of nuclear matter. Incompressibility is given as its
second derivative with respect to baryon density, at the
saturation point

\begin{equation}
K=9\left[\rho^{2}\frac{\partial^{2}}{\partial\rho^{2}}
(\frac{\varepsilon}{\rho_{B}})
\right]_{\rho=\rho_{0}} \;.
\end{equation}

Symmetry energy of nuclear matter is defined as a second
derivative of binding energy per nucleon with respect to the
asymmetry parameter
$\alpha=(\rho_{p}-\rho_{n})/(\rho_{p}+\rho_{n})$:

\begin{equation}
\varepsilon(\rho,\alpha)=\varepsilon(\rho,0)
+S_{2}(\rho)\alpha^{2}+S_{4}(\rho)\alpha^{4},
\end{equation}
where parameters $S_{2}, S_{4}$ are defined as

\label{symmetryparams}
\begin{eqnarray}
S_{2}=\frac{1}{2}\left[\frac{\partial^{2}\varepsilon(\rho,\alpha)}
{\partial\alpha^{2}}\right]_{\alpha=0}
\;,
\\
S_{4}=\frac{1}{24}\left[\frac{\partial^{4}\varepsilon(\rho,\alpha)}
{\partial\alpha^{4}}\right]_{\alpha=0}
\;.
\end{eqnarray}

Parameter $S_{2}$ is often used as symmetry energy itself,
argumented by negligible contribution of higher order $S_{4}$
parameter, especially for densities relevant for common nuclei.

\section{Results and discussion}

The mean-field parametrizations were obtained by calculation using
three different DBHF results for nuclear matter as initial data -
results of Li, Brockmann and Machleidt \cite{BLM92}, results of
Lee, Kuo, Li and Brown \cite{LKLB98,LKLB97} and finally
calculations of Huber, Weber and Weigel \cite{HWW93,HWW95}.

Based on the realistic and relativistic NN interaction of the Bonn
group, in the work \cite{BLM92} there were performed DBHF
calculations that yield an effective NN interaction, and
subsequently the single-particle potentials, equations of state,
nucleon effective masses, and speed of sound for both symmetric
and neutron matter were studied. The Bonn A potential reproduced
quantitatively the empirical saturation properties of nuclear
matter as well as nucleon effective mass.

The work \cite{LKLB98} deals with asymmetric matter also using the
DBHF approach with Bonn A one-boson-exchange NN interaction. Not
only saturation properties, but in addition even the empirical
value of the symmetry energy at the saturation density were
reproduced satisfactorily. Isoscalar meson potentials for
symmetric matter authors calculated in \cite{LKLB97}.

Finally energy per nucleon for several asymmetries using DBHF
approach is calculated in \cite{HWW93}, together with proton and
neutron scalar and vector potentials from \cite{HWW95} making
possible to fit also the mean-field parameter for coupling of
$\delta$ meson to nucleons. Furthermore, in this two works it is
tested also momentum dependence of self-energies, however, due to
momentum independence of mean-field theory coupling constants, the
momentum independent results were chosen to fit.

As it is easily seen from equations for meson potentials and for
energy per nucleon, the squares of coupling constants appear
exclusively in ratios with meson masses and thus one can fix the
meson masses to experimental values without any physical
restriction of the RMF. The meson masses considered in this work
are $m_{\sigma}=550$ MeV, $m_{\omega}=783$ MeV, $m_{\rho}=770$ MeV
and $m_{\delta}=980$ MeV.

\begin{table}
\caption{\label{table1}Parameter sets resulting from the RMF fit
to DBHF results of Machleidt {\sl et al.} \cite{BLM92}.}
\begin{ruledtabular}
\begin{tabular}{ccc}

 & \mbox{MA} &
  \mbox{MB} \\

\hline

$g_{\sigma}^{2}$ & 106.85 & 112.27 \\

$g_{\omega}^{2}$ & 180.61 & 204.36 \\

$g_{\rho}^{2}$ & 18.445 & 9.4932 \\

$b_{\sigma}$ & -0.0025823 & -0.0029820 \\

$c_{\sigma}$ & 0.011529 & 0.013345 \\

$c_{\omega}$ & 0.015849 & 0.020449 \\

$\Lambda_{V}$ & 0.25857 & --- \\

$\chi^{2}/N$  & 2.76  & 9.95 \\

\end{tabular}
\end{ruledtabular}
\end{table}

\begin{figure}
\includegraphics[width =7cm]{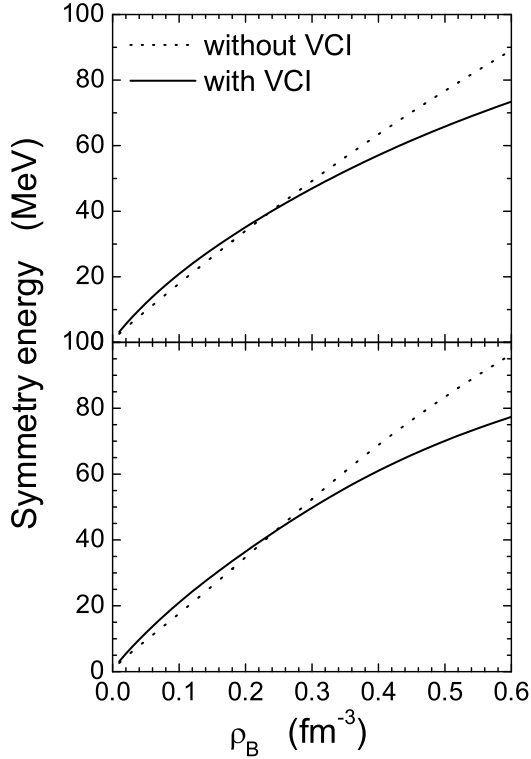}
\caption{\label{figure0} Density dependence of the symmetry energy
for two different parametrization sets. Upper panel shows results
for the fit of Machleidt {\sl et al.} DBHF results \cite{BLM92},
with mean-field parametrizations listed in Table~\ref{table1},
where $\sigma$, $\omega$ mesons with their selfinteractions and
$\rho$ mesons was used as degrees of freedom, and both with and
without inclusion of vector meson cross-interactions (VCI),
plotted with solid and dotted lines, respectively. The lower panel
displays analogical calculation results for the fit of Lee {\sl et
al.} DBHF results \cite{LKLB98,LKLB97}, with parametrizations in
Table~\ref{table2}.}
\end{figure}

The first fit was performed using $\sigma$, $\omega$, $\rho$
mesons and scalar isoscalar cubic and quartic selfinteractions,
vector isoscalar quartic selfinteractions as well as
cross-interactions between vector mesons, fitting energies per
nucleon for symmetric and neutron matter and symmetric isoscalar
potentials. The corresponding parameter set obtained is listed in
Table~\ref{table1} - for comparison with (parametrization MA) and
without (MB) vector meson cross interaction. This parametrization
reproduces the DBHF results satisfactorily in whole fitting range
of densities relevant for common nuclei, which can be said both
for energy per nucleon and also for isoscalar $\sigma$ and
$\omega$ meson potentials. Growth of the $\rho$ potential is
decreasing with baryon density, being result of the isovector
cross-interactions. This has impact on the energy per nucleon
especially for extreme asymmetries and consequently on curvature
of symmetry energy, which is plotted in the Fig~\ref{figure0},
upper panel. The cross-interactions significantly affect density
dependence of the symmetry energy - they increase rise of symmetry
energy below 0.24 fm$^{-3}$ and slower it above this density.
Symmetry energy at the saturation density is 33.3 MeV, which is in
accordance with experimental value of about 34 MeV. The
incompressibility of symmetric matter at the saturation density is
347 MeV.

For better description of asymmetry behavior of matter it is of
course profitable to use not only symmetric and neutron case, but
also other nonsymmetric cases with partial fraction of protons, as
was calculated with DBHF theory by Lee {\sl et al.}
\cite{LKLB98,LKLB97}. Results of the fit with the same degrees of
freedom as were in the previous case, performed for binding energy
per nucleon for several asymmetries as well as for symmetric
matter scalar and vector potentials, are listed in
Table~\ref{table2} (parametrization LA), also with parameter set
obtained using model without vector meson cross-interactions (LB).
All of the relevant physical quantities for all of the asymmetries
are reproduced closely. Also in this parametrizations there is
strong influence of cross-interaction, which is displayed in
Fig~\ref{figure0}, lower panel. There is increasing growth of
symmetry energy below approximately 0.25 fm$^{-3}$ and
deceleration in higher density region. This fact will be comment
more closely in the last data set.

\begin{table}
\caption{\label{table2}Parameter sets resulting from the fit of
Lee {\sl et al.} DBHF calculations \cite{LKLB98,LKLB97}.}

\begin{ruledtabular}
\begin{tabular}{ccc}

& LA & LB \\

\hline

$g_{\sigma}^{2}$ & 103.91  & 102.11  \\

$g_{\omega}^{2}$ & 147.84  & 146.73  \\

$g_{\rho}^{2}$ & 17.432 & 9.6697 \\

$b_{\sigma}$ & 0.00097186  & 0.00083559 \\

$c_{\sigma}$ & 0.0012694 & 0.0012411 \\

$c_{\omega}$ & 0.0054204 & 0.0051878 \\

$\Lambda_{V}$ & 0.18790 &  ---   \\

$\chi^{2}/N$  & 1.69 & 2.62  \\

\end{tabular}
\end{ruledtabular}
\end{table}

\begin{table}
\caption{\label{table3}Parameter sets resulting from the fit of
Huber {\sl et al.} DBHF results \cite{HWW93,HWW95}.}

\begin{ruledtabular}
\begin{tabular}{ccccc}

& HA &
 HB & HC & HD \\

\hline

$g_{\sigma}^{2}$ & 90.532  & 86.432 & 91.110 & 87.591 \\

$g_{\omega}^{2}$ & 108.95  & 106.89 & 109.26 & 107.61 \\

$g_{\rho}^{2}$ & 36.681 & 28.795 & 20.804 & 15.335 \\

$g_{\delta^{2}}$ &  28.739 & 25.170 & --- & --- \\

$b_{\sigma}$ & 0.0043852  & 0.0033779 & 0.0044388 & 0.0035745 \\

$c_{\sigma}$ & -0.0052045 & -0.0037762 & -0.0052076 & -0.0039753 \\

$c_{\omega}$ & -0.0001421 & -0.0010509 & -0.0000385 & -0.0007753 \\

$\Lambda_{V}$ & 0.10647 &  --- & 0.34805 & --- \\

$\chi^{2}/N$  & 2.05 & 3.80 & 5.85 & 6.89 \\

\end{tabular}
\end{ruledtabular}
\end{table}

\begin{figure}
\includegraphics[width =7cm]{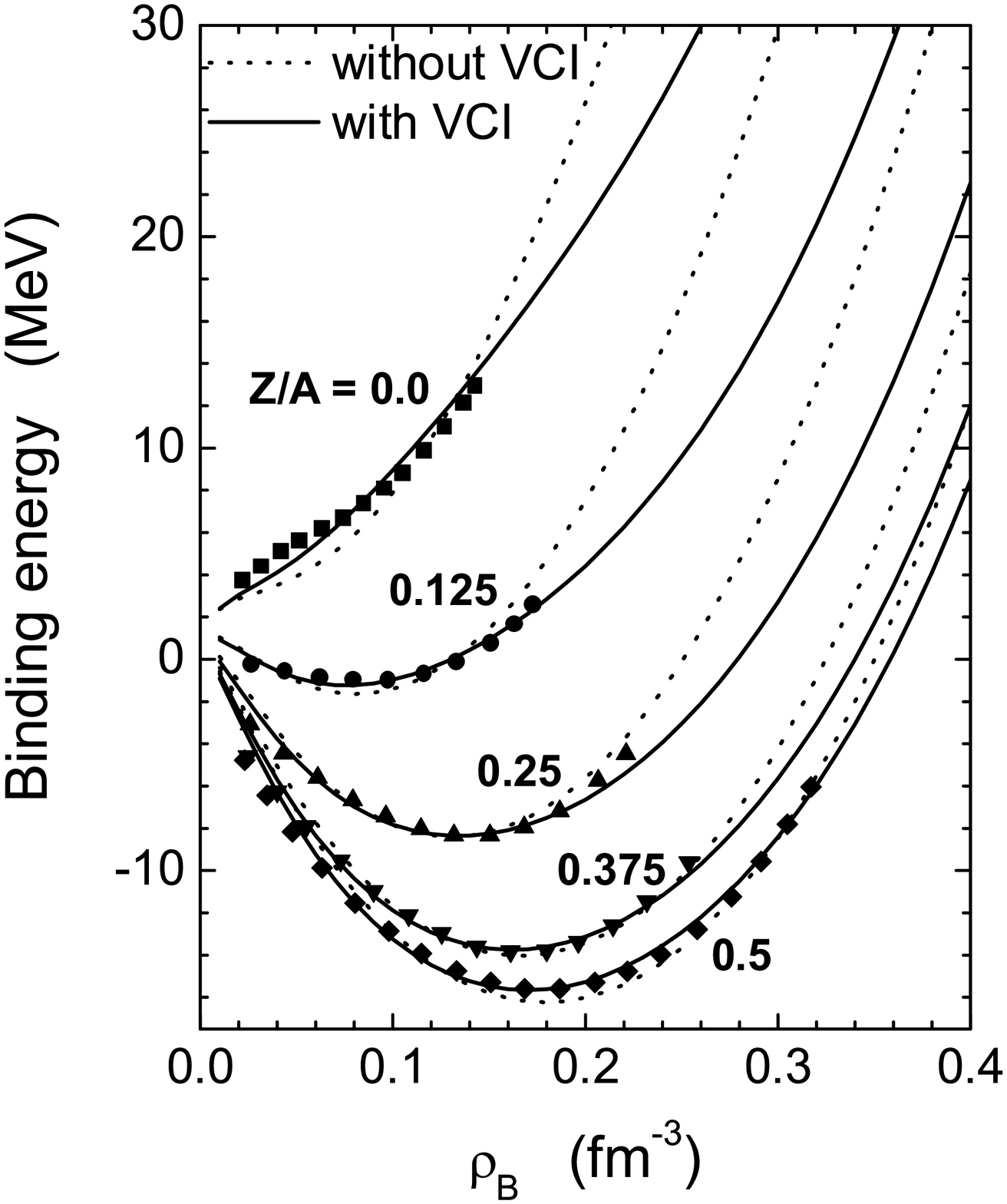}
\caption{\label{figure1}Density dependence of the binding energy
per nucleon for five different asymmetries from pure neutron
matter to symmetric nuclear matter, resulting from mean-field
theory fit (represented by lines) of Huber {\sl et al.} DBHF
results \cite{HWW93,HWW95} (scatter symbols), with corresponding
parametrizations listed in Table~\ref{table3}. Isoscalar $\sigma$
and $\omega$ mesons with their selfinteraction, isovector $\rho$
mesons with (solid lines, parametrization HA) and without (dotted
lines, parametrization HB) cross-interaction with $\omega$ mesons,
and $\delta$ mesons was used as degrees of freedom.}
\vspace{0.5cm}
\end{figure}

In the theoretical framework we was dealing also with
isovector-scalar sector of nucleon-nucleon effective interaction.
In the work \cite{HWW93} there was calculated binding energy per
nucleon for several asymmetries in the DBHF approach, using Bonn
potential B with density independent self-energies. Additionally
in \cite{HWW95} authors performed calculation also for proton and
neutron scalar and vector potentials, enabling us to fit not only
$\rho$ potential but also $\delta$ potential value. The fit was
thus using $\sigma$, $\omega$ mesons, their selfinteractions,
$\rho$ meson with its cross-interaction to $\omega$ meson, and
finally even $\delta$ meson as degree of freedom. Results of the
fit for binding energy per nucleon for several considered
asymmetries are drawn in the Fig~\ref{figure1}. Fit values
(represented by lines) follow closely the DBHF results (scatter
symbols) for all asymmetries. Corresponding parameter set is
listed in Table~\ref{table3} (HA), together with fit parameters
obtained without vector cross-interactions (HB).

\begin{figure}
\includegraphics[width =7cm]{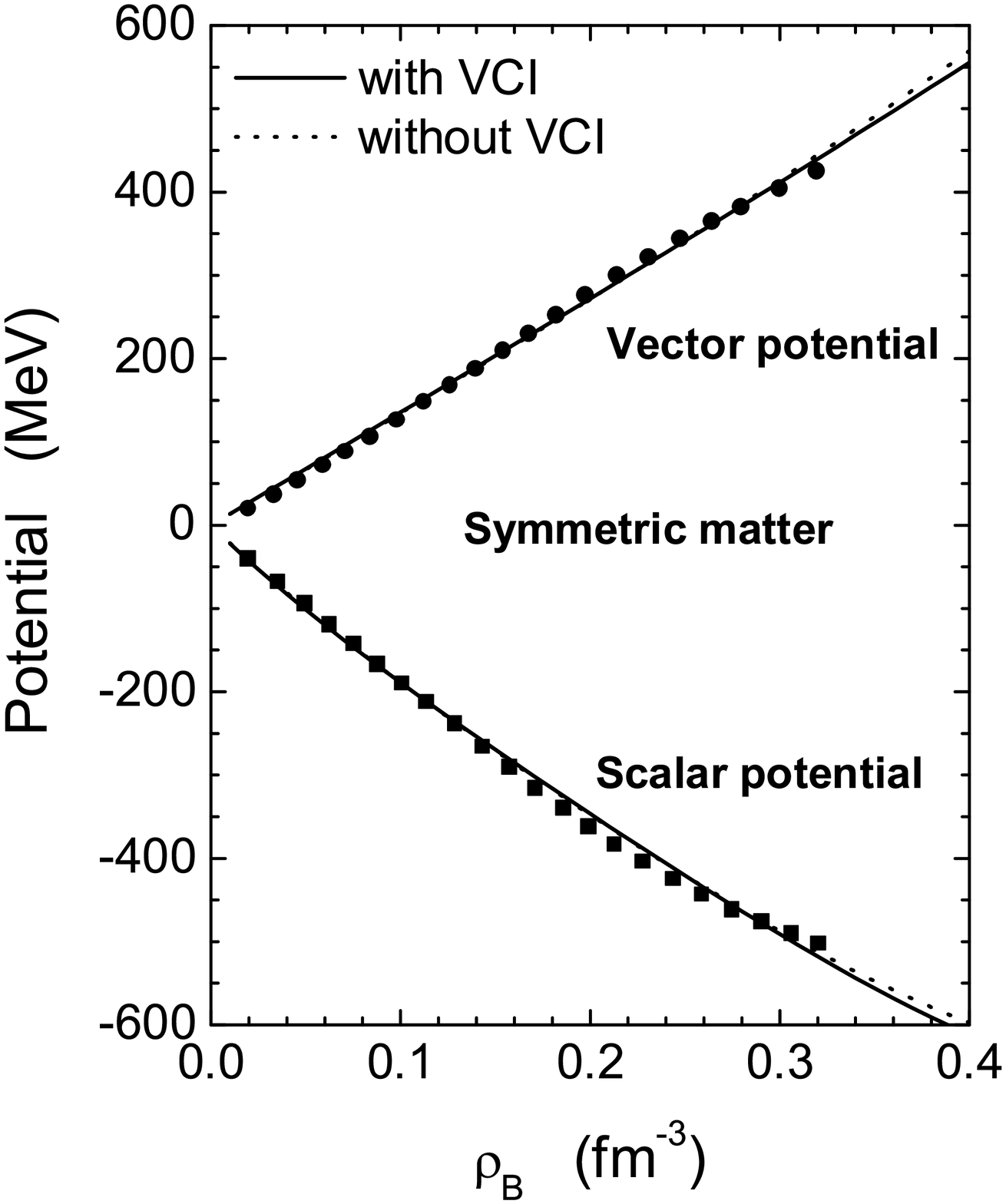}
\caption{\label{figure2}Isoscalar potentials of symmetric nuclear
matter, resulting from the fit of Huber {\sl et al.} DBHF results
\cite{HWW93,HWW95}, with the same parametrization and notation as
in the Fig~\ref{figure1}.}
\end{figure}

Negative values of the quartic selfinteraction constants are
consequence of the fact, that dependence of symmetric scalar and
vector potentials ensued from DBHF calculations on the baryon
density is almost linear, which brings some difficulties into the
determination of the selfinteraction force of isoscalar mesons,
and thus results in negative and positive second derivative of
vector and scalar potential, respectively. One of the explanations
of this verity could be the fact, that the Van-Hove theorem
\cite{HVH58} is, unlike for the DBHF theory, fully consistent with
the mean-field model only. The fit of Lee {\sl et al.} DBHF
calculations is not the case (see Table~\ref{table2} - all
selfinteraction constants have positive sign), due to more
distinct curvature of density dependence of the potentials (not
shown in this paper). The incompressibility of nuclear matter is
235 MeV.

\begin{figure}
\includegraphics[width =7cm]{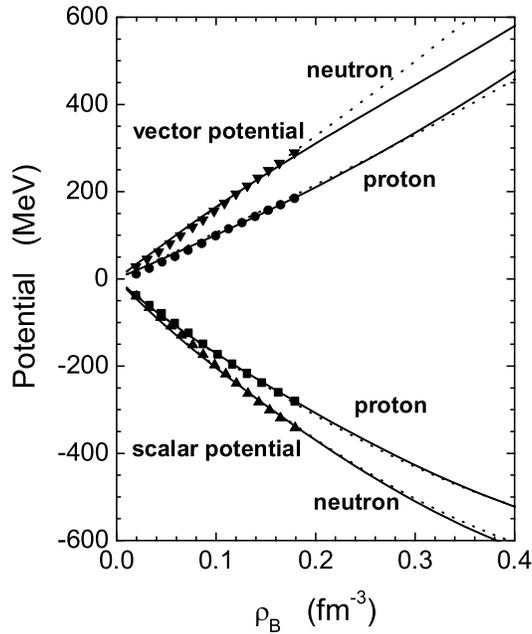}
\caption{\label{figure3}Scalar and vector potentials of protons
and neutrons of the asymmetric matter with proton fraction Z/A =
0.125, for the parametrization and notation identical with
previous figures.} \vspace{0.5cm}
\end{figure}

Simultaneously with binding energies fit to the symmetric scalar
and vector potentials (Fig~\ref{figure2}) and also to
above-mentioned proton and neutron scalar and vector potentials
for proton fraction 0.125 (Fig~\ref{figure3}) was performed. As
can be seen all of these potentials are reproduced satisfactorily
within several MeV.

\begin{figure}
\includegraphics[width =7cm]{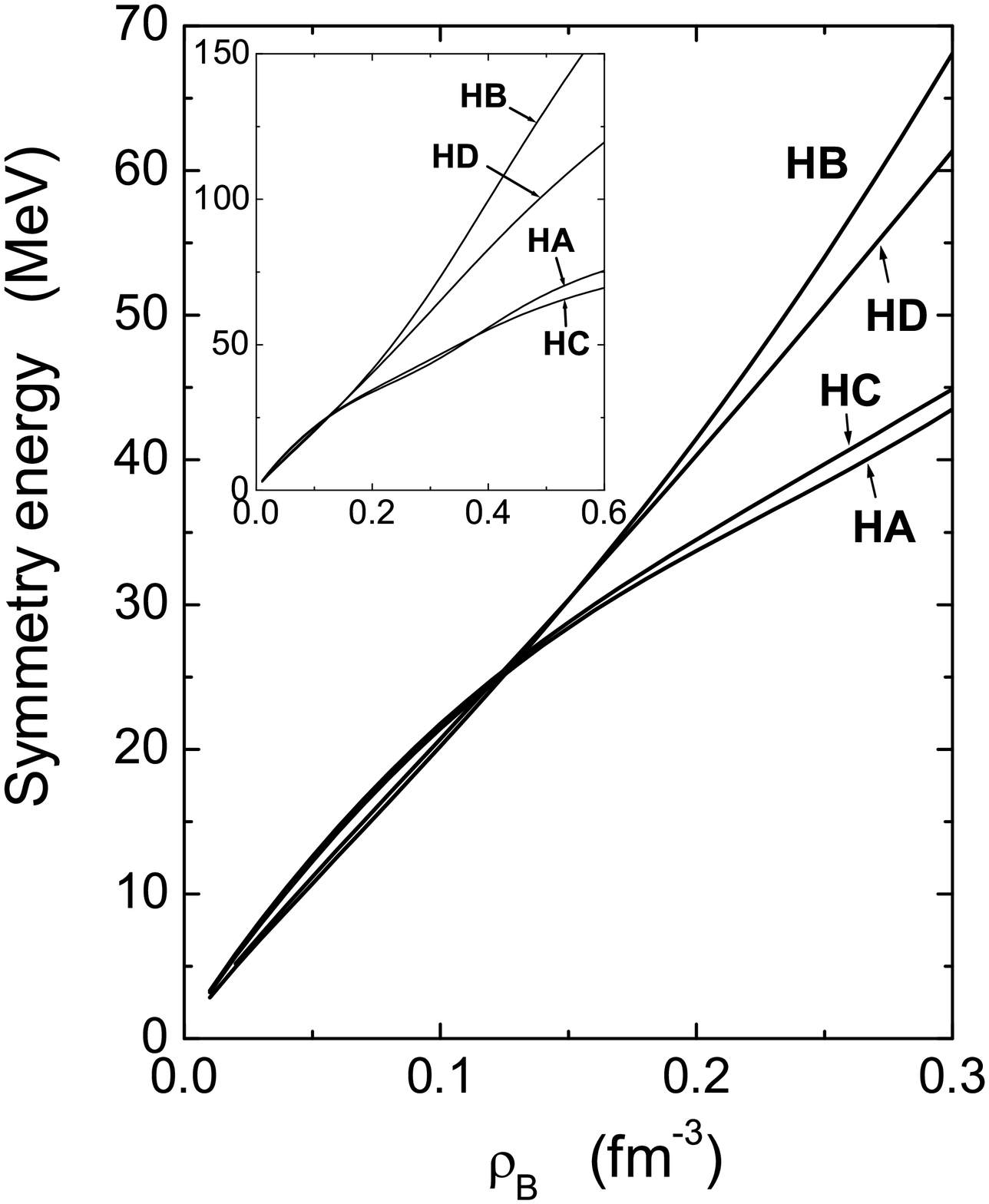}
\caption{\label{figure4}Density dependence of the symmetry energy
of nuclear matter, as resulted from the fit of Huber {\sl et al.}
DBHF results \cite{HWW93,HWW95} for several considered degrees of
freedom (all models include also isoscalar scalar and vector
selfinteractions, see Table~\ref{table3}): HA:
$\sigma$,$\omega$,$\rho$ and $\delta$ mesons and $\rho$-$\omega$
cross-interaction; HB: $\sigma$,$\omega$,$\rho$ and $\delta$
mesons; HC: $\sigma$,$\omega$,$\rho$ mesons and $\rho$-$\omega$
cross-interaction; HD: $\sigma$,$\omega$ and $\rho$ mesons. The
inset picture shows the symmetry energy behavior in a wider
density region.}
\end{figure}

Evaluation of $\rho$-$\omega$ cross-interaction influence on
symmetry energy and its comparison with $\delta$ meson influence
is possible from Fig~\ref{figure4}. There it is drawn density
dependence of the symmetry energy, where each of the lines is
correspondent to different degrees of freedom used (see
Table~\ref{table3}): HA contains both cross-interactions and
$\delta$ meson contribution; then inclusion of the $\delta$ meson
without vector cross-interaction is accomplished by HB; HC
considers the vector cross-interaction but without $\delta$ meson,
and finally basic model with $\sigma$ and $\omega$ mesons with
their selfinteractions and $\rho$ mesons only (HD). Vector
cross-interaction has significant impact on symmetry energy, where
its influence has analogical nature as in the previous two data
sets, thus supporting those results also in the case of $\delta$
meson inclusion. It increases growth of the symmetry energy in
this case below approximately 0.13 fm$^{-3}$, which is important
for description of properties of exotic nuclei near the dripline
and is similar as consequences of density dependence of the
isovector couplings due to Fock contributions, which was
calculated in \cite{GCTF01}. Above 0.13 fm$^{-3}$ it slowers
symmetry energy rise thus implicating an impact on higher density
behavior of matter (high energy beam collisions, supernovae
explosions, neutron stars properties), and it is also in agreement
with recent Hartree calculations \cite{HP01}, where similar
high-density influence of cross-interactions was concluded. In
comparison $\delta$ meson contributes relatively slightly to
behavior of symmetry energy - for lower and intermediate densities
its contribution is almost inconspicuous thus not significantly
affecting properties of atomic nuclei. For higher densities in
absence of cross-interaction it slightly increases symmetry
energy. Also this result is in concordance with recent
calculations of other authors \cite{GBC02}. However, the presence
of cross-interaction results beside strong curvature of symmetry
energy also in contrary (compared to previous case) effect of
$\delta$ meson - for higher densities it softly fortifies decrease
of symmetry energy growth. This leads to conclusion, that
$\rho$-$\omega$ cross-interactions seems to be an important degree
of freedom which should be used in further calculations.

\section{Summary}

In this work the relativistic mean field theory was used to obtain
an effective parametrization of the properties of asymmetric
nuclear matter calculated by more fundamental
Dirac-Brueckner-Hartree-Fock theory. The energy per nucleon
together with the symmetric isoscalar potentials were fitted, and
simultaneously also proton and neutron scalar and vector
potentials. Isoscalar $\sigma$, $\omega$ mesons with their
selfinteractions, and isovector $\rho$, $\delta$ mesons with
$\rho$-$\omega$ cross-interaction were used as degrees of freedom
and parameters of the fit. Generally a good reproduction of both
the energy and the potentials was reached and thus the parameter
sets are representing an effective DBHF description of asymmetric
nuclear matter at normal baryon densities applicable as well for
calculation of finite nuclei properties. The cross-interaction
between $\rho$ and $\omega$ mesons turns out to improve
reproduction of properties of asymmetric nuclear matter and
additionally it increases symmetry energy in common nuclei density
region, e.g. approximately bellow saturation density, and
decreases this energy above saturation point, thus having
consequences for properties of finite nuclei, especially with
large isospin asymmetry, and also for description of nuclear
matter at higher densities, relevant in high energy nuclear
collisions and several astrophysical processes and phenomena (e.g.
neutron star properties and supernovae explosions). Isovector
$\delta$ meson also improves quality of the mean-field model, but
without such a strong impact on density dependence of symmetry
energy. This results imply that $\rho-\omega$ cross-interaction is
very useful for better description of nuclear matter and that it
could be important for calculations of properties of finite nuclei
with high isospin asymmetry.

\begin{acknowledgments}
This work was supported by the Slovak Grant Agency for Science
VEGA under grant N$\sharp$ 2/1132/21. Calculations were performed
at the Computational Centre of Slovak Academy of Sciences.
\end{acknowledgments}

\end{document}